\documentclass{iopart}
\usepackage{sistyle}
\usepackage{graphicx}
\usepackage{hyperref}
\begin{document}
\author{Robert Spero$^1$,  Brian Bachman$^1$, Glenn de Vine$^1$, Jeffrey Dickson$^1$, 
William Klipstein$^1$, Tetsuo Ozawa$^1$, Kirk McKenzie$^1$,
 Daniel Shaddock$^{1, 2}$  David Robison$^1$,
 Andrew Sutton$^2$ and Brent Ware$^1$ }
\address{$^1$Jet Propulsion Laboratory (JPL), California Institute of Technology, 4800 Oak Grove Drive Pasadena, CA 91109 USA}
\address{$^2$Centre for Gravitational Physics, The Australian National University, ACT 0200 Australia}
\begin{abstract}
Recent advances at JPL in experimentation and design for LISA  interferometry include the demonstration of Time Delay Interferometry  using electronically separated end stations, a new arm-locking design with improved gain and stability, and progress in flight readiness of digital and analog electronics for phase measurements.
\end{abstract}
\title{Progress in Interferometry for LISA at JPL}
\date{8 February, 2011} 
\submitto{\CQG}

\maketitle

\section{Frequency noise then and now}
  In describing the sensitivity of LISA to laser frequency noise \cite{tinto-etc}, it is often asserted that without Time Delay Interferometry (TDI), the strain sensitivity as limited by frequency noise would be equal to the laser's fractional frequency noise multiplied by the fractional arm-length mismatch:  
  \begin{equation}
  \label{michelson}
   \tilde{h}(f) = (\tilde{\nu}(f)/\nu_o)(\Delta L/L),
   \end{equation}
   where $\tilde{\nu}(f)$ is the square root of the power spectral density of the frequency noise at the output of a laser with average frequency $\nu_0,$ $L$ is the interferometer arm-length, and $\Delta L$ is the arm-length mismatch.
 Typically, $\tilde{\nu}(f)/\nu_o =(\SI{30}{Hz/\sqrt{Hz}})/(\SI{3e14}{Hz})=\SI{1e-13}{/\sqrt{Hz}},$ 
 $\Delta L/L=(\SI{5e7}{m})/(\SI{5e9}{m})=0.01,$ yielding $\tilde{h}(f)=\SI{1e-15}{/\sqrt{Hz}},$ several orders of magnitude larger than the  sensitivity after applying TDI.  This analysis is for a Michelson interferometer, in which the light from a single laser is split between two arms, then interferometrically recombined. For such detectors, it is indeed beneficial to match the arm lengths, $\Delta L\approx 0.$  
 
LISA doesn't work that way.  Instead, two independent lasers are interfered for each length measurement, and the Michelson interferometer signal is synthesized from  two length measurements.  
 TDI is not just a method for reducing frequency noise, it is the prescription for combining measurements that individually have much more power in the noise than in the signal.  The concept of frequency noise sensitivity before the application of TDI has little meaning.  Instead of Equation~\ref{michelson},
 the frequency noise sensitivity in a TDI measurement is 
 \begin{equation}
  \tilde{h}(f) = (\tilde{\nu}(f)/\nu_o)(c\Delta \tau/L),
  \end{equation}
   where $c$ is the speed of light and $\Delta\tau$ is the error in the {\em knowledge} of the time synchronization between measurements.  Loosely,  $\Delta l=c\Delta\tau$ is the uncertainty in the separation, or ``range'' between proof masses. The measurements that determine $\Delta \tau,$ however, are all directly related to timing, not distance---primarily the error in synchronizing clocks on separated spacecraft.  These timing measurements are insensitive to $L$ or $\Delta L$.  Consequently, adjusting the orbit to minimize $\Delta L$ has little effect on frequency noise.  Also, TDI experiments add little of relevance by incorporating large LISA-like delays, $\tau=L/c=\SI{17}{s}.$  The experiments we describe emphasize $\Delta\tau$ and $\tilde{\nu}(f)$, not $\tau.$
 
Another misconception on the limits to LISA performance in the presence of frequency noise relates to the role of coherence length $L_{\rm coh}$ of the laser output.  Associated quantities are coherence time $\tau_{\rm coh}=L_{\rm coh}/c$ and linewidth $\Delta\nu = 1/(\pi t_{\rm coh}).$   $L_{\rm coh}$ is defined as the allowable length mismatch before interference degrades significantly.  For $\Delta\nu =\SI{30}{Hz},$ $L_{\rm coh}=\SI{3e6}{m}$, smaller than the typical $\Delta L$ in LISA.   One might conclude that LISA operates with significantly degraded interference.  In fact, the frequency noise even from lasers with  $\Delta\nu\gg\SI{30}{Hz}$ does not  degrade the interference signal.  The LISA instrument measures the  phase between interfering lasers, and the coherence length requirement is replaced by the requirement that the relative phase fluctuations $\Delta\phi$ not exceed the capability of the measurement apparatus (plus the requirement that the beat note frequency stay within the $\sim\SI{20}{MHz}$ bandwidth of the phasemeter).  The measurement is tolerant of up to \SI{1}{cycle} of fluctuation in phase within \SI{10}{\mu s};  larger fluctuations will cause phasemeter cycle-slipping.  This robust capability allows measuring interference between completely independent, unstabilized lasers exhibiting noise similar to that of non-planar ring oscillator (NPRO) lasers.
  
Our understanding of how to keep laser frequency noise out of the LISA
science signal has matured rapidly in recent years.  This stands in
contrast to other aspects of the LISA design that have not suffered
fundamental change since the ``LISA Pre-Phase A Report, Second Edition" (PPA2) of 1998~\cite{PPA2}.
The general structure of the mission described in PPA2, as well as many
of the details, remain intact:  the gravitational wave sensitivity, orbital
configuration,  gravitational reference sensor, thrusters, clocks, and laser hardware of today's
baseline design are all similar to the original design.  Not so for the
measurement and control of laser frequency noise, even at the conceptual
level:  
\begin{itemize}
\item The original phasemeter was  based on zero-crossing detection,
100\,kHz beat frequencies, and tracking filters---a design that would
introduce unacceptable noise from aliasing and other imperfections
absent in the current phasemeter design~\cite{phasemeter}, and that is incompatible with multi-tone measurements.
\item The current baseline method of correcting the influence of frequency noise
in phase measurements according to  individually measured
arm lengths does appear in PPA2, though with different parameters
($\Delta l=\SI{200}{m}$ resolution then vs. as small as \SI{1}{cm} now).  But the concepts of
decimation, interpolation and TDI for LISA
data  streams, and the general approach of forming different signal combinations as part of ground-based post-processing,
 did not exist in 1998.  Rather, a frequency-domain method
of unspecified bandwidth was described;  we now know that that method would
be impracticable to implement.  
\item Arm-locking was incorrectly believed to be limited in bandwidth by the
long travel times between the arms, and hence was not included in the
design.
\end{itemize}

These advances in aggregate---TDI, wide-dynamic range decimating
phasemeters, interpolation, and high-gain arm-locking---have
transformed the problem of frequency noise from nearly intractable to
one with robust solutions offering orders of magnitude of performance
margin~\cite{white-paper}. 

 \section{Time Delay Interferometry (TDI) Experiment}
 The most effective tool in the LISA design suite for suppressing laser frequency noise is Time Delay Interferometry.  The concept and a recent experimental test are described in \cite{jpl-tdi}.  We review here some of the experimental details and  the main results.
 
\begin{figure} 
\begin{center}
\includegraphics[width=9cm]{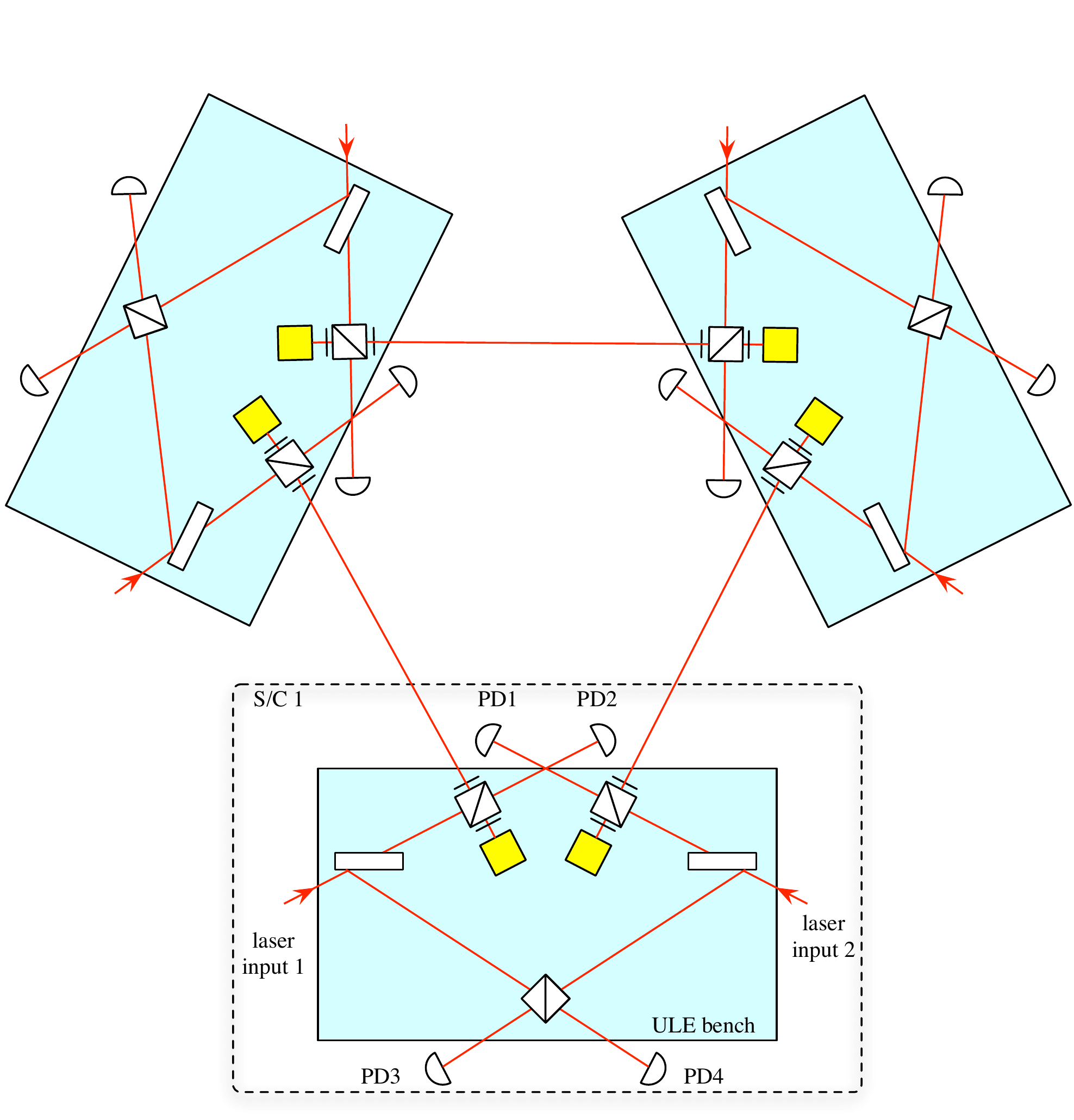}\\[4mm]
\includegraphics[width=16cm]{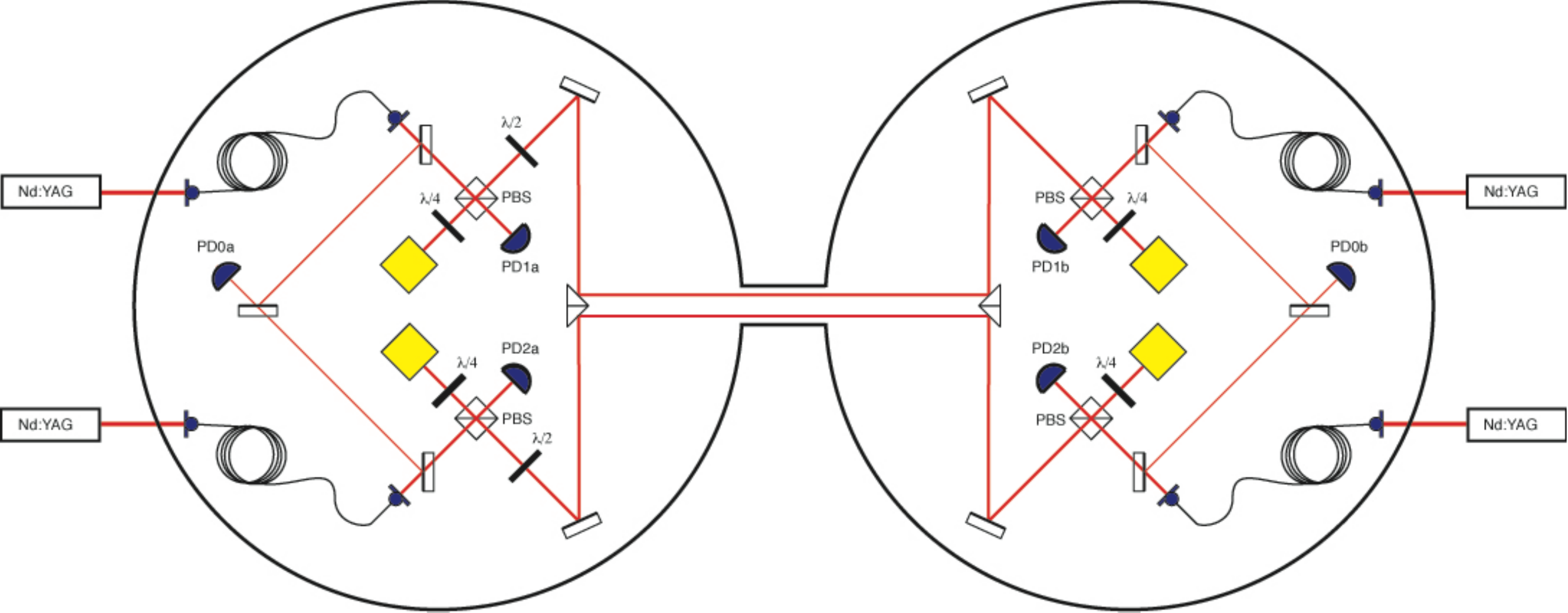}
\caption{Beam paths in the LISA spacecraft configuration (upper) and in the TDI laboratory experiment (lower).  PD = photodetector, PBS = polarizing beamsplitter, $\lambda/4$ = quarter-wave plate, Nd:YAG = neodymium-doped yttrium aluminium garnet laser.\label{beam-paths}}
\end{center}
\end{figure}

The beam paths for the LISA interferometer and in the TDI experiment are indicated in Figure~\ref{beam-paths}.  In the experiment, two vacuum chambers stand in for two of the three LISA spacecraft.  The omission of the third spacecraft is equivalent to operating that spacecraft as an optical transponder.  In common with the spacecraft configuration, associated with each chamber are two frequency offset phase-locked lasers, with typical frequency difference \SI{4}{MHz}. Each laser is split to provide a local oscillator, via a small amount of light that passes through a crossed polarizer, as well as the measurement beam directed to the other chamber.   The local oscillator beams interfere with each other to provide the signal for offset-locking.   The incoming beams also interfere with the local oscillators;  as with LISA, the local oscillator power is much larger than the incoming power, to provide a strong signal against noise in the photoreceiver.  The lasers are free-space NPRO Nd:YAG type, and provide approximately \SI{100}{mW} power at  \SI{1064}{nm} wavelength.  Optical fibers within the vacuum chambers reject pointing fluctuations from these free-space lasers.  The laser beams enter the chambers through windows, and the interfered beams also pass through windows to extra-chamber photoreceivers.  The phase measurements are recorded at a rate of \SI{100}{Samples/s} for high-accuracy interpolation in post-processing.

\begin{figure} 
 \begin{center}
  \includegraphics[width=15cm]{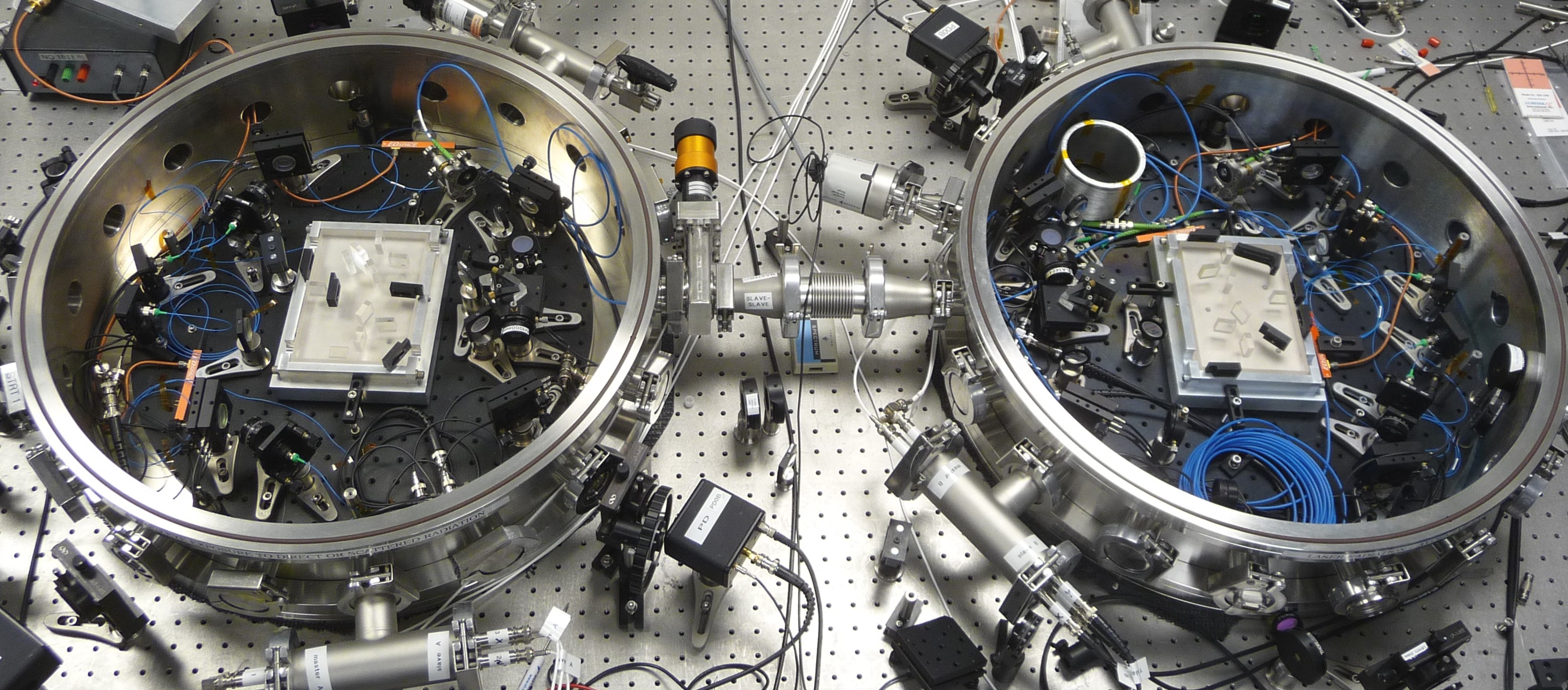}
   \caption{Vacuum chambers for the TDI experiment.  Path-sensitive optical elements are optically contacted to ULE plates in the chamber centers.\label{photo}}
   \end{center}
 \end{figure}

Path-sensitive optical elements are mounted on  rectangular  Ultra Low Expansion (ULE)\cite{corning} plates, visible in the chamber centers in
Figure~\ref{photo}.  
Maintaining sensitivity to optical path lengths while demonstrating cancellation of laser frequency noise
is key to exposing non-linear effects, an attribute of this TDI
demonstration absent in similar work~\cite{mitryk}.
Each vacuum chamber also holds electro-optic modulators used for transfer and subsequent elimination of clock noise, and for time coordination and optical data communication~\cite{sutton}.  Clock noise is eliminated by modulating sidebands spaced approximately \SI{6}{GHz} from the central optical frequency, and \SI{1}{MHz}  pseudo-random noise (PRN) code provides the other functions.  The laser control and interferometer readout for each chamber is connected to a set of electronics independent of the electronics at the other end.   Clock synchronization, in particular, is provided by PRN modulation as opposed to an electrical connection.

\begin{figure} 
\begin{center}
\includegraphics[width=4cm]{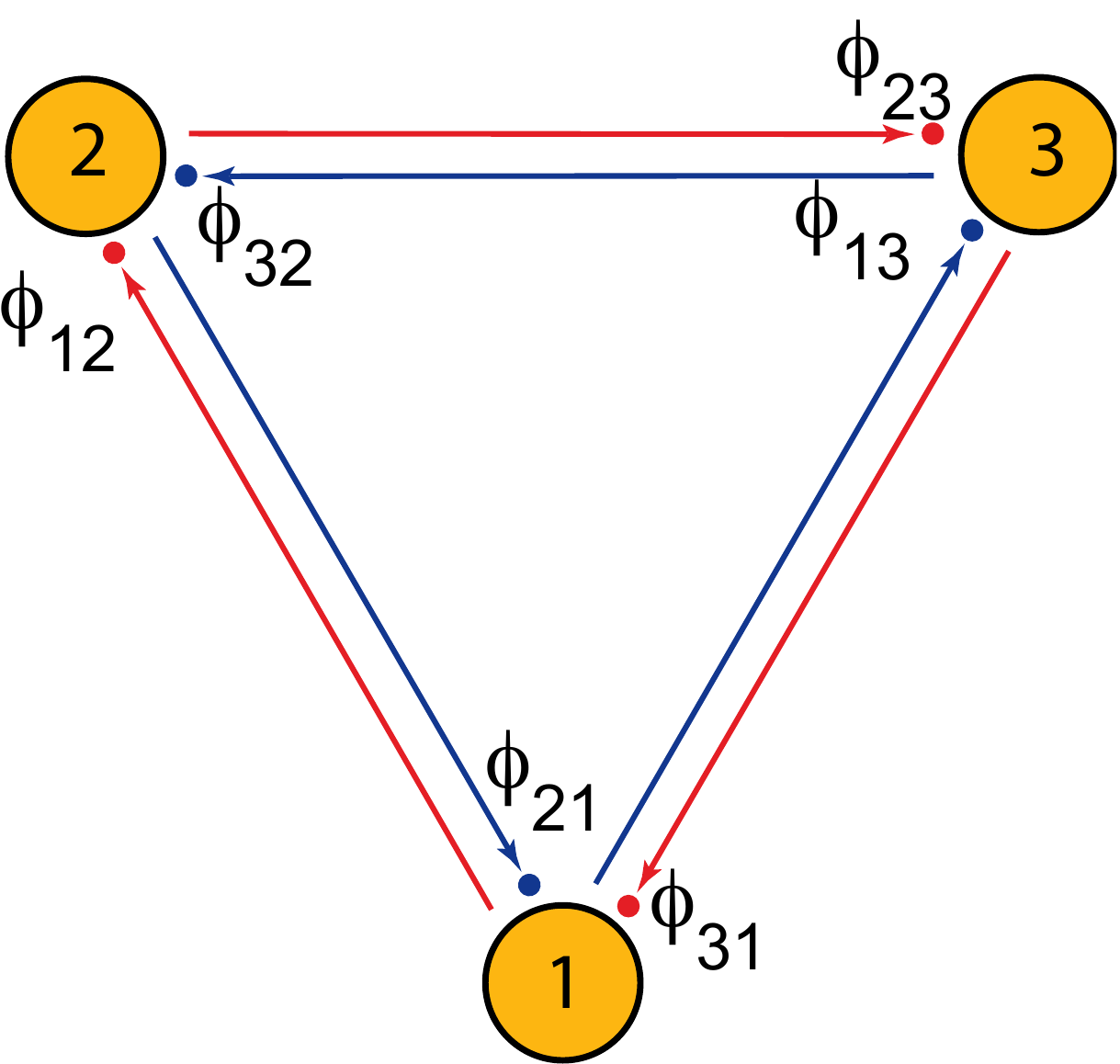}
\caption{Sagnac configuration, showing three spacecraft and six phase measurements.\label{sagnac-diag}}
\end{center}
\end{figure}

Figure~\ref{sagnac-diag} shows schematically the beam paths and phase measurements for the TDI Sagnac combination~\cite{shaddock-sagnac}.  The sensitivity of this combination to gravitational waves is similar to that of a Michelson interferometer.  We selected the Sagnac combination because it is insensitive to the paths between vacuum chambers, as the counter-propagating beams overlap.  It retains full sensitivity to motion of the mirrors that stand in for proof masses,  and to the dynamics of laser frequency and clock noise.  Also, its  requirements on timing synchronization are similar to those that apply to all LISA TDI signal combinations.

Each small dot at the arrowheads in Figure~\ref{sagnac-diag} represents a phase measurement between a local laser and an incoming beam.  For example, $\phi_{13}$  is the phase difference between the incoming beam from spacecraft\,3 and the local laser at spacecraft\,1.  The Sagnac TDI combination $\alpha$ is equivalent to the phase measured by a conventional Sagnac interferometer:

\begin{eqnarray}
\label{alpha-3}
\alpha(t)&=&\phi_{31}-\phi_{21}
+D_{23}D_{31}\phi_{12}-D_{21}\phi_{32}
-D_{32}D_{21}\phi_{13}+D_{31}\phi_{23}.
\end{eqnarray}
The delay operators are defined by $D_{ij} a(t) = a(t-L_{ij}/c),$ where $L_{ij}$ is the path length for a beam traveling from $i$ to $j$.  Designating the phase of the laser beam originating at spacecraft $i$ and traveling toward spacecraft $j$ as $s_{ji},$ the phase measurements are
\begin{eqnarray}
\phi_{ij}&=&D_{ij}s_{ij}(t)-s_{ij}(t).
\end{eqnarray}
Equation~\ref{alpha-3} is simplified, in that it omits the measurement of the phase between local oscillators and the timing synchronization measurements that correct for clock offsets.
For the two-station laboratory TDI experiment, Equation~\ref{alpha-3} further simplifies to
\begin{eqnarray}
\label{alpha-2}
\alpha(t)&=&\phi_{31}-\phi_{21}
+D_{23}D_{31}\phi_{12}-D_{21}\phi_{32}.
\end{eqnarray}

\begin{figure} 
\begin{center}
\includegraphics[width=10cm]{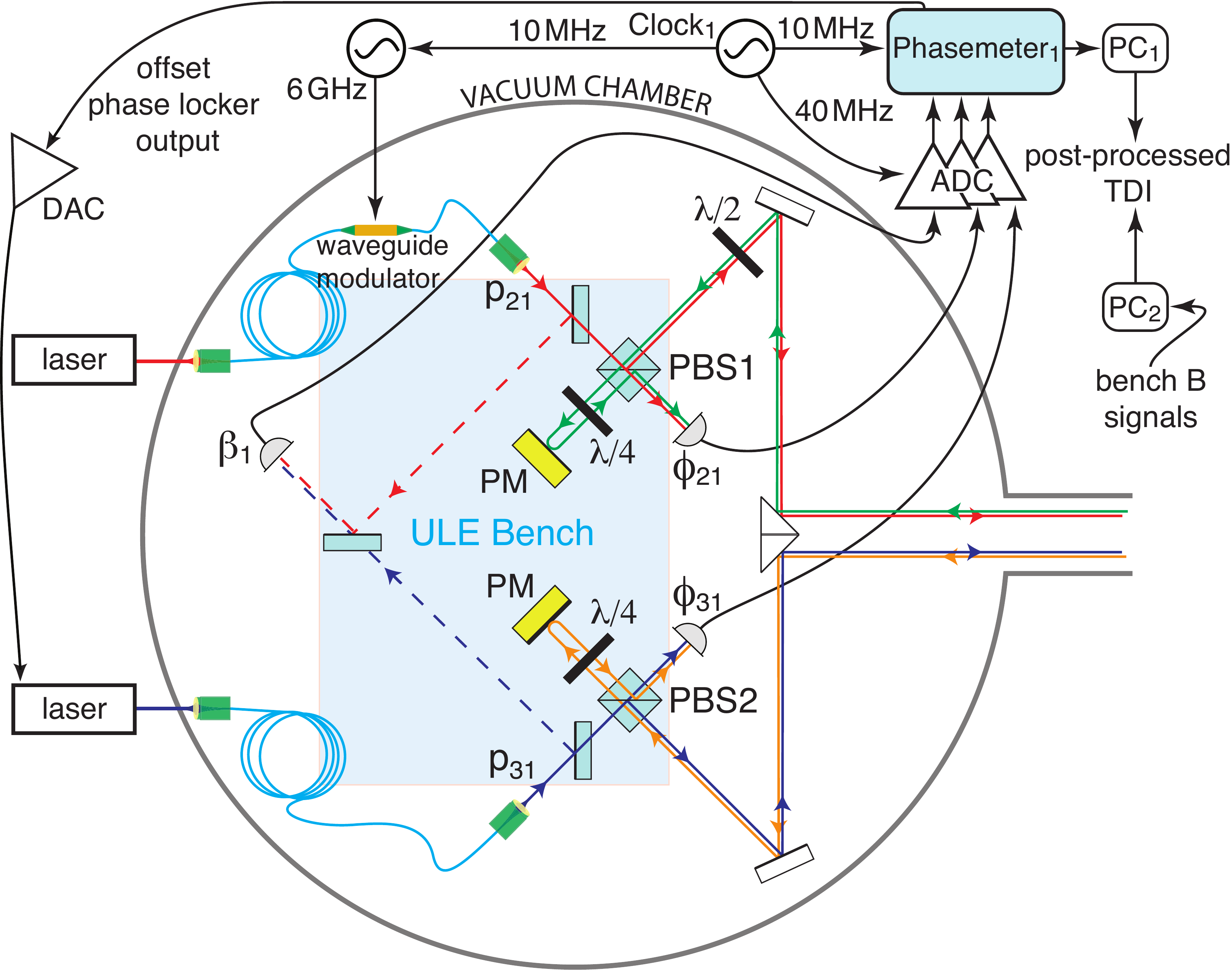}
\caption{Optics and electronics associated with one of the vacuum chambers.\label{single-chamber}}
\end{center}
\end{figure}

The error terms in $\alpha(t)$, Equation~\ref{alpha-2}, are of the form
$ \alpha_e=\tau_e\,d\phi/dt$
where $\tau_e$ is a (ns-scale) error in delay estimation, and $d\phi/dt$ is a (MHz-scale) interference frequency.  
The resulting spectral density noise is
\begin{equation}
\label{alpha_e}
\tilde{\alpha}_e=\tau_e\frac{d\tilde{\phi}}{dt}+\tilde{\tau}_e\frac{d\phi}{dt}
\end{equation}
Here $\tau_e$ is the difference of two different errors:  (range error) - (timing offset).
Uncorrected, this error is large:  for clock fractional frequency error $\Delta y = \SI{1e-6}{}$, $\tau_e=(\Delta y)T,$ giving $\tau_e=\SI{1}{ms}$ after $T=\SI{1000}{s}.$
The first term in Equation~(\ref{alpha_e}) is the error from frequency noise and range/offset error; the
second term  is due to the phase noise in the clocks,  is independent of laser frequency noise, and is proportional to the heterodyne frequency.
The two terms are corrected by two separate subsystems:  MHz-scale PRN timing synchronization and GHz-scale clock noise correction, respectively.
Figure~\ref{single-chamber} shows the  waveguide modulators used for both types of measurement.

\begin{figure} 
\begin{center}
\includegraphics[width=10 cm]{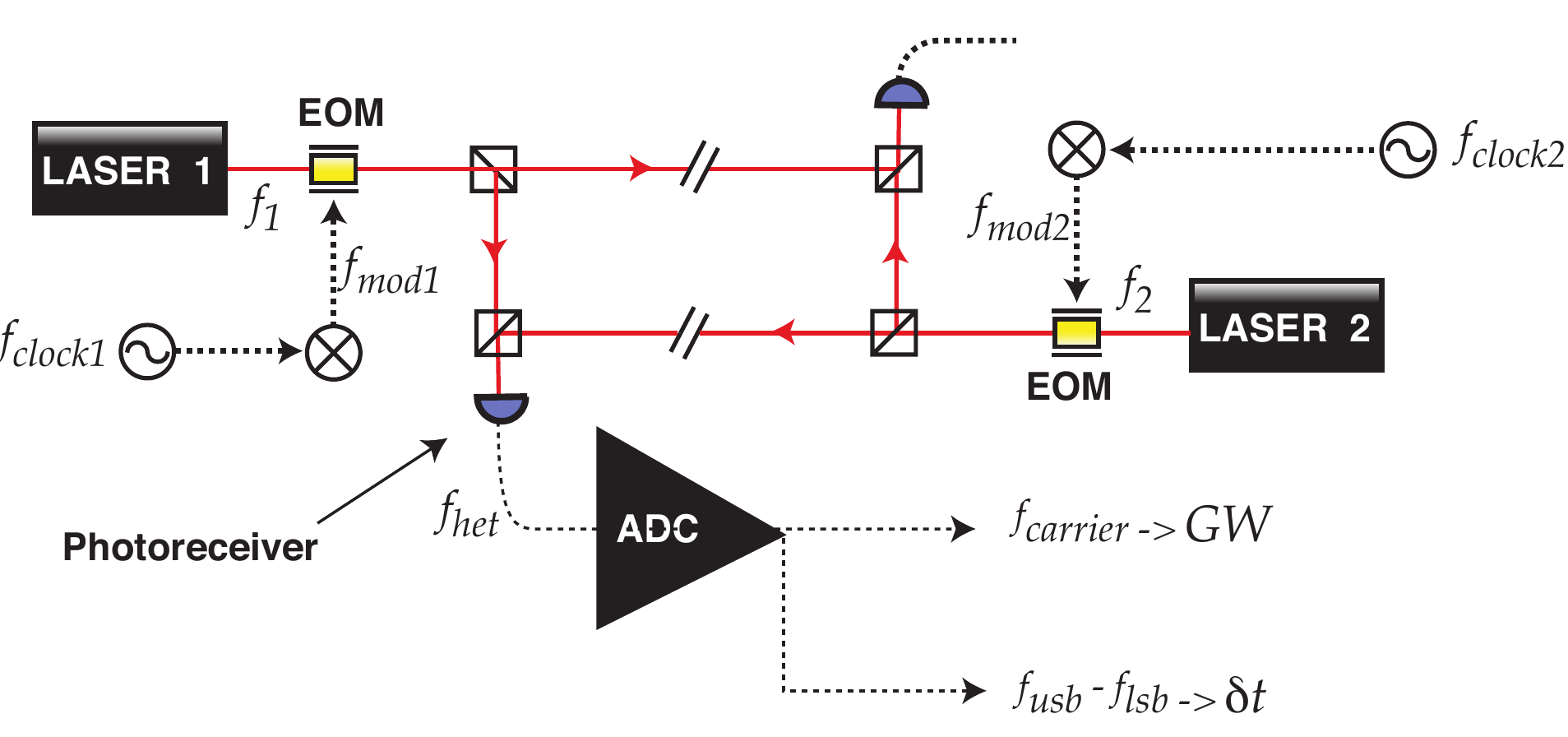}
\includegraphics[width=2.5cm]{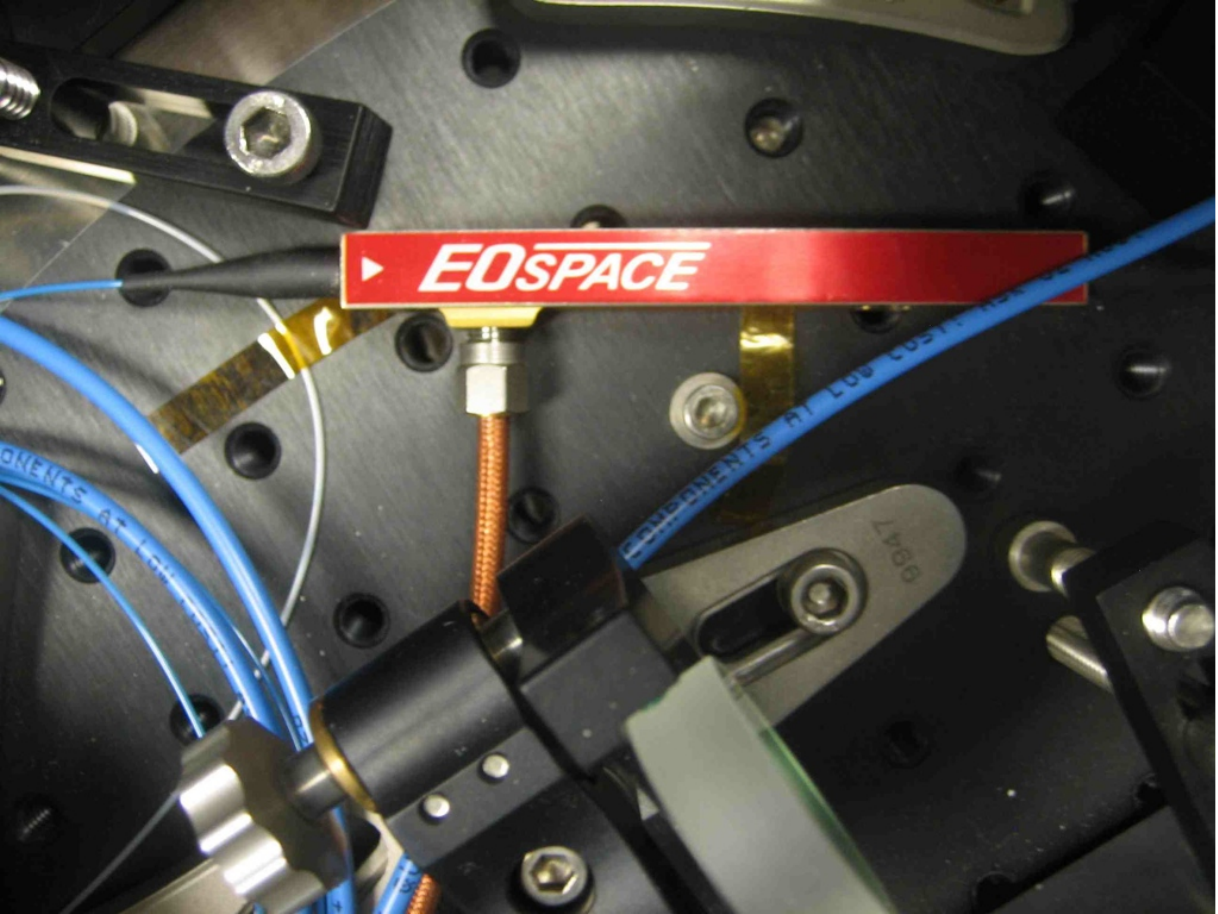}
\caption{Modulation and demodulation used for clock noise correction, left, and photograph of electro-optic modulator in vacuum chamber, right.
\label{clock-correction}}
\end{center}
\end{figure}
 The clock frequency is multiplied to 6\,GHz for high signal-to-noise measurement  of clock phase fluctuations,
transmitted to the opposite station as sidebands on the science signal,
 and detected as a sideband/sideband beat signal at frequency $\sim \SI{2}{MHz}$.
This clock tone is separated from the science signal within the phasemeter and recorded along with the main science phase measurement.  The modulation and demodulation are illustrated in Figure~\ref{clock-correction}.  The measurement of timing synchronization by PRN codes~\cite{sutton} is not shown.

The photoreceiver signals, such as $\phi_{21}$ in Figure~\ref{single-chamber} are digitized by analog to digital converters (ADC) and processed by field-programmable gate array (FPGA) based electronics.  Measurements of some key properties of the ADC/FPGA combination are listed in Table~\ref{pm-table}.  These measurements meet or exceed the requirements for phase measurement in LISA.
\begin{table}
\begin{center}
\begin{tabular}{|c|c|}\hline

Property & Measurement limit \\ \hline\hline
Nonlinearity&$<\SI{e-14}{}$\\ \hline
$2f_0$ suppression & $>\SI{70}{dB}$\\ \hline
Ampltude feedthrough $\frac{d\phi}{d\alpha}$& $<\SI{e-6}{cycle}$\\ \hline
Quantization noise & $<\SI{e-7}{cycle/\sqrt{Hz}}$\\ \hline
Frequency slew rate & $>\SI{7e5}{Hz/s}$\\ \hline
Frequency range&$\SI{2-18}{MHz}$\\ \hline
Weak-light  acquisition & $<\SI{40}{pW}$ \\ \hline
\end{tabular}
\caption{Key properties of the JPL phasemeter.\label{pm-table}}
\end{center}
\end{table}
 The results reported here were obtained with a variety of single-element commercial photoreceivers, using optical signal powers greater than \SI{10}{\mu W}, compared to the design power for LISA of \SI{200}{pW}.  Consequently, photon shot noise does not contribute significantly to the observed noise.

\begin{figure} 
\begin{center}
   \includegraphics[width=10cm]{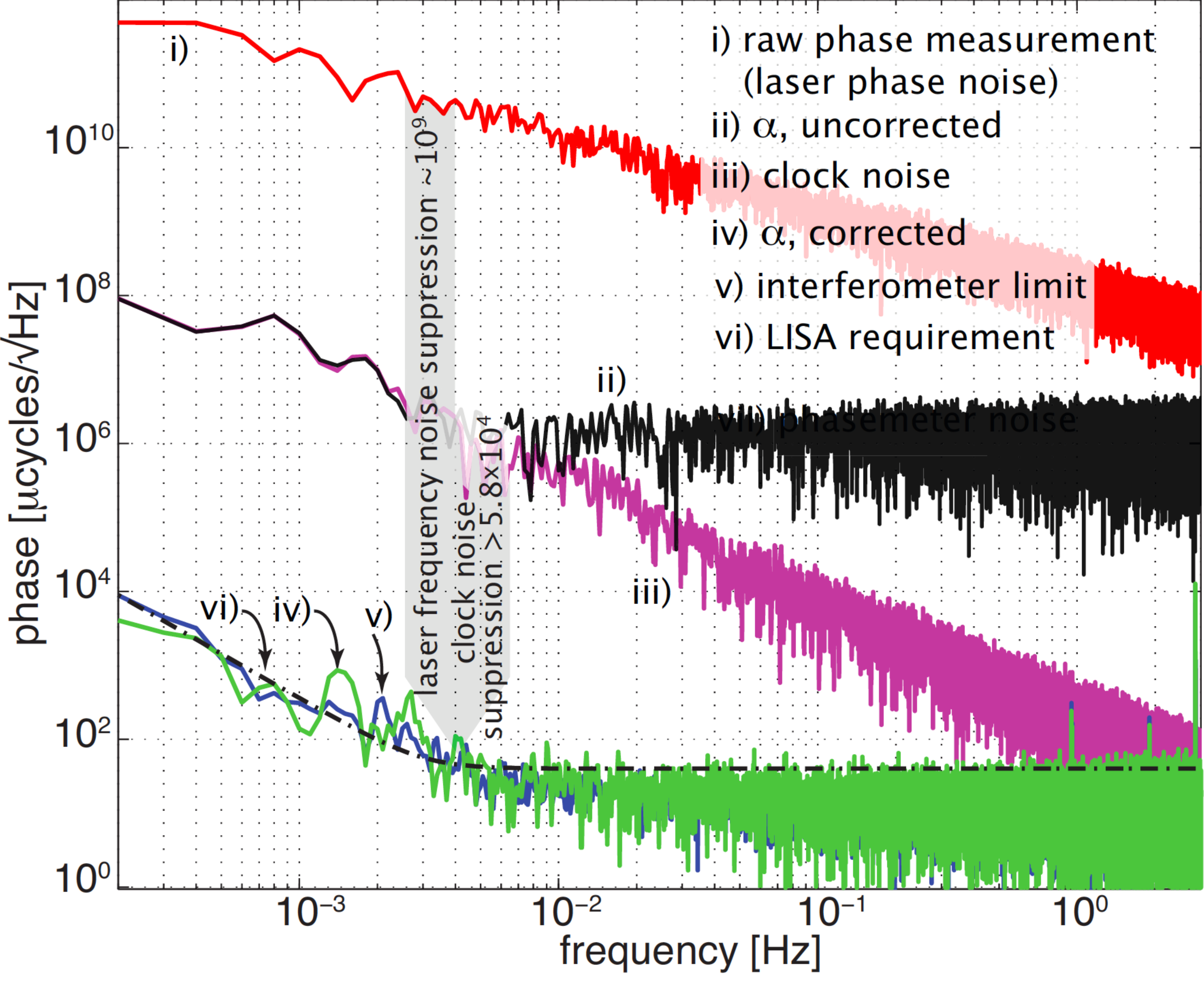}q
   \caption{Noise spectra from the TDI experiment.\label{tdi-results}}
   \end{center}
   \end{figure}

Results from the TDI experiment are shown in the noise spectra, Figure~\ref{tdi-results}. 
The uppermost trace (i) shows laser frequency noise, amplitude 
  $\SI{800}{Hz/\sqrt{Hz}}$ injected in the offset-locking loop.  The injected noise tests the dynamic range of the measurement similarly to the situation in LISA if the master laser has $\SI{800}{Hz/\sqrt{Hz}}$ after stabilization to a reference cavity or other frequency reference.
Curve (iii) shows the laser frequency noise removed after high-accuracy interpolation and the application of TDI, Equation~\ref{alpha-2}.  After the removal of clock noise using the interpolated clock sidebands, the final sensitivity is
 given by curve (iv). 
This level  matches the displacement noise limit of the interferometer, measured with phase-locked clocks and no injected laser frequency noise.
The measured suppression of laser frequency noise, a factor of \SI{1e9}{}, is within a factor of 10 of the LISA requirement for the contribution to the total error budget from frequency noise, and the
suppression of clock noise, factor of \SI{6e4}{}, far exceeds the LISA requirement of 10--1000.

\begin{figure}  
\begin{center}
\includegraphics[width=7cm]{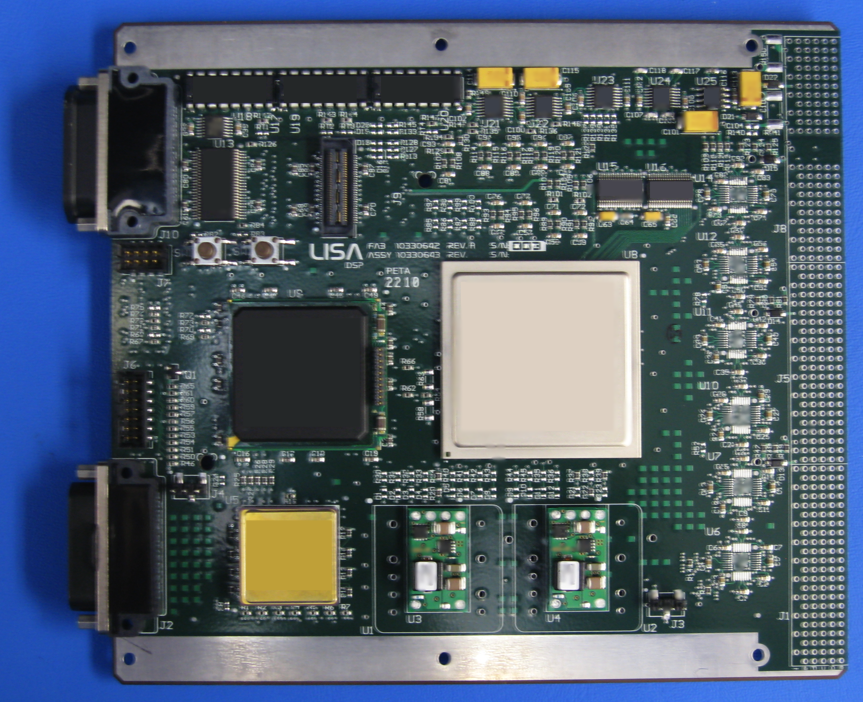}\qquad
\includegraphics[width=6cm]{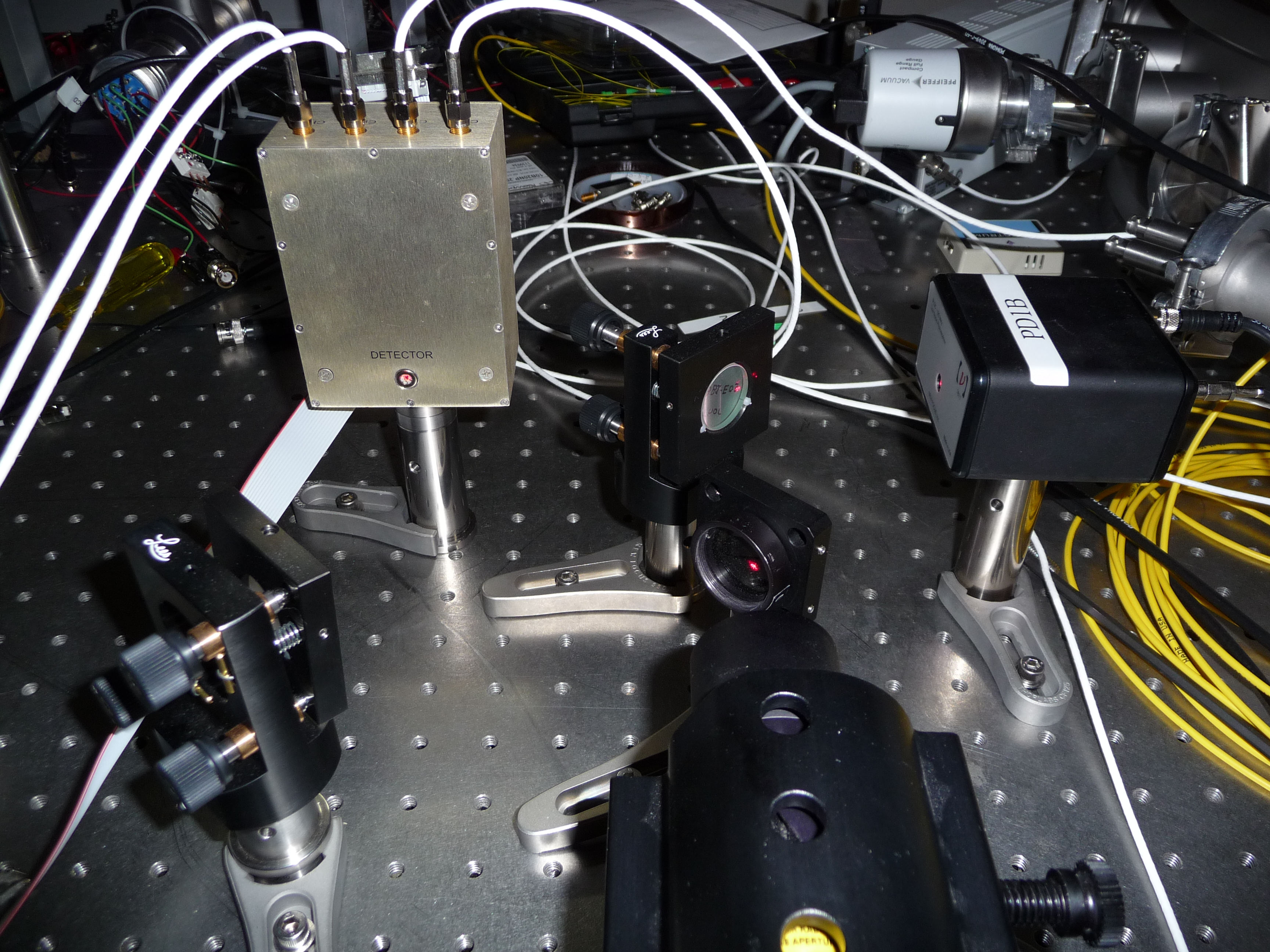}
\caption{Digital signal processing board, left, and quadrant photoreceiver, right.\label{flight-hw}}
\end{center}
\end{figure}

Concurrently with improvements to the laboratory experiment, we are also developing hardware for flight, Figure~\ref{flight-hw}.  A prototype digital signal processing board is at technology readiness level (TRL) 5 , meaning it is built with flight-like hardware.  
We are also developing quadrant photoreceivers appropriate for science measurements and for wavefront sensing;  TRL 4 versions 	have a noise-equivalent power (NEP) of $\rm {NEP}<\SI{5}{pW/\sqrt{Hz}}$ in the LISA frequency band, and meet the requirements on frequency-dependence of group delay. 

\section{Arm Locking}
\begin{figure} 
\begin{center}
\includegraphics[width=10cm]{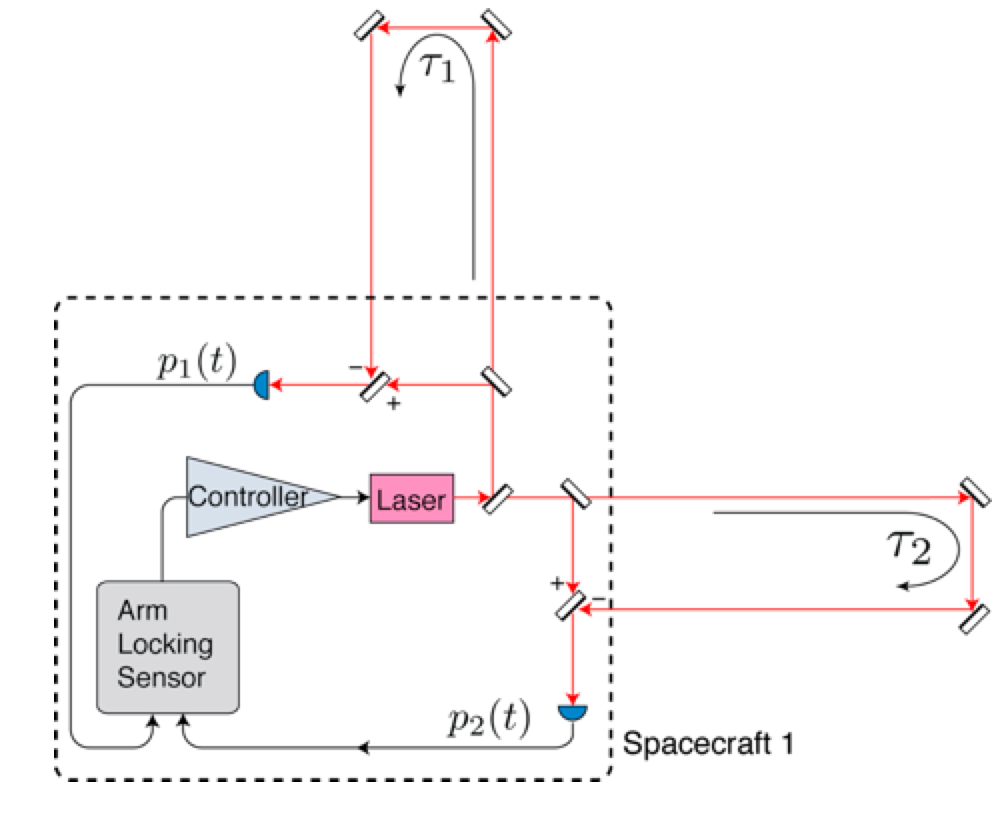}
\caption{Arm-locking concept.\label{arm-concept}}
\end{center}
\end{figure}

\begin{figure} 
\begin{center}
\includegraphics[width=7cm]{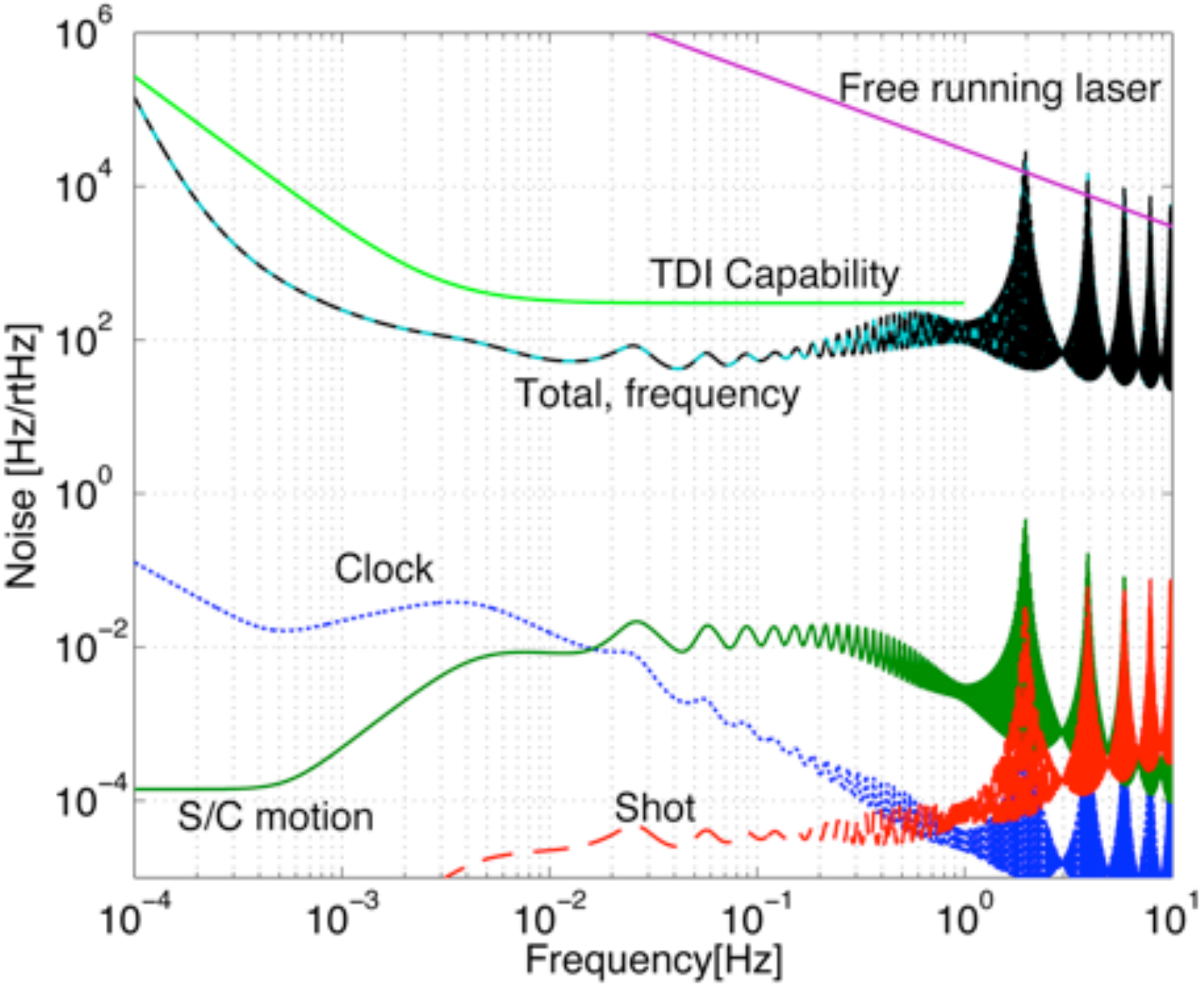}\qquad
\includegraphics[width=7cm]{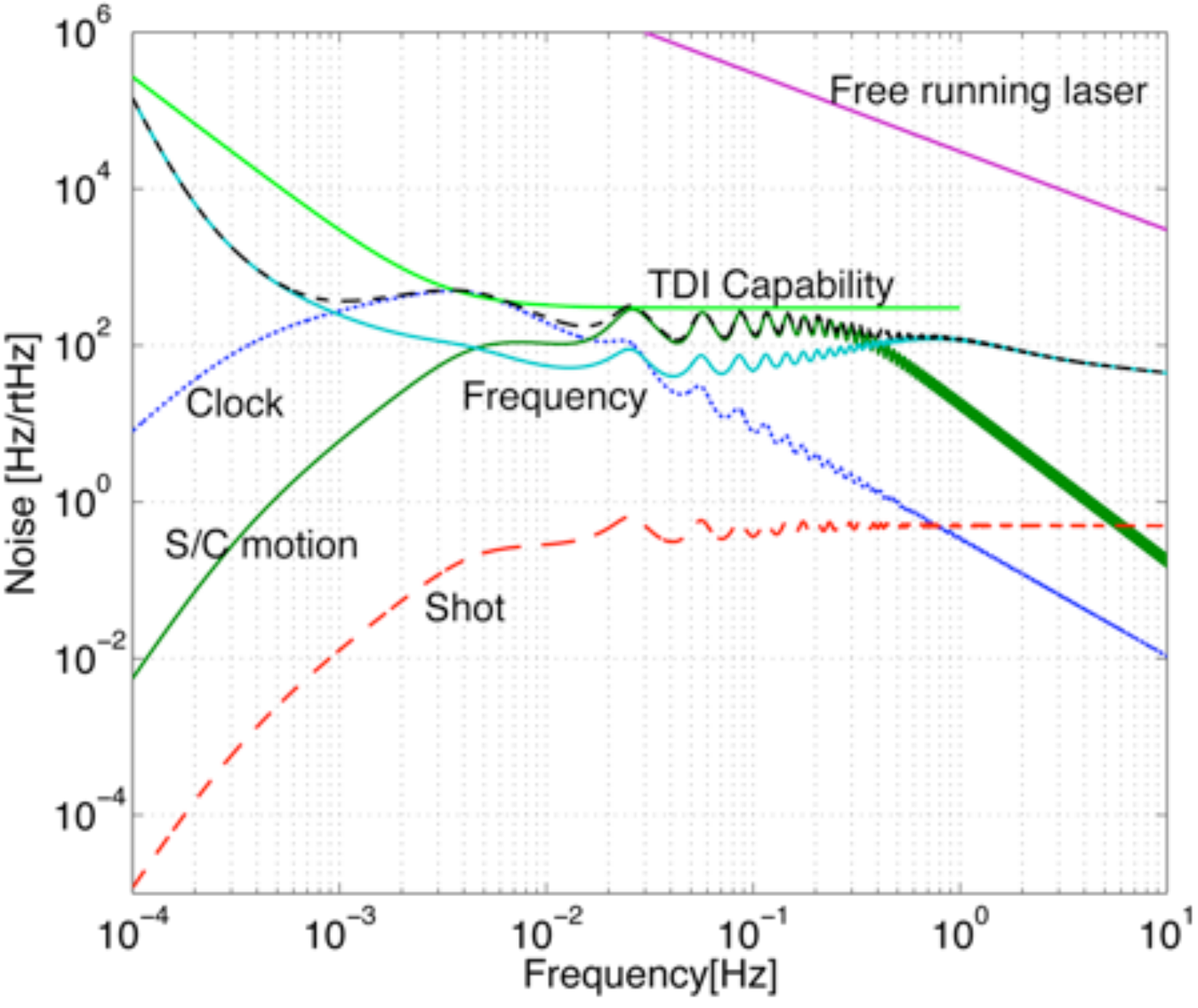}
\caption{Arm-locking with $\Delta L = \SI{7.5e7}{m},$ gain-limited (left) and with
$\Delta L = \SI{1.2e4}{m},$ noise-limited (right).\label{al-perf}.}
\end{center}
\end{figure}

Arm locking is a technique for stabilizing the LISA  master laser frequency using the very stable separation between proof masses in one or more of the LISA arms.  The concept using two arms is illustrated in Figure~\ref{arm-concept}.  The interference between the prompt and  delayed light from the laser is measured as $p_{1,2}$ for round-trip arm delays $\tau_{1,2}.$   These measurements are combined to form the arm-locking sensor, and filtered by the controller to stabilize the laser.  A recent improvement in the technique is based on the modified dual arm locking sensor~\cite{dual-al}.
With an optimized controller, the 
upper unity gain frequency can be made as high as \SI{15}{kHz}.
Frequency pulling effects are reduced by high-pass filtering in the controller, giving a low-frequency unity-gain frequency of $\SI{5}{\mu Hz}.$  Frequency  noise after arm locking, starting with a free-running NPRO laser, is illustrated in Figure~\ref{al-perf}.   The two extremes of arm length mismatch $\Delta L$  are shown.  In both cases and for all $\Delta L$ values in between, the arm-locking performance meets the LISA requirements for frequency noise, without the need for other reference or stabilization systems.

\section{Conclusion}
In 2010, the ``Astro 2010'' decadal survey report of the
National Academy of Sciences recommended\cite{astro2010} that the LISA mission be given
a high priority among large astrophysics space missions.  The work reported here---experimental demonstration of frequency noise cancellation by TDI, development of electronic hardware suitable for flight, and improved pre-TDI stabilization using arm locking---advance the vision of that recommendation, an orbiting gravitational wave detector returning signals to earth in the next decade.

This research was performed at the Jet Propulsion Laboratory, California Institute of technology, under contract with the National Aeronautics and Space Administration, with support from Australian Research CouncilÕs Discovery Projects funding
scheme (project number DP0986003).

\end{document}